# State-Based Behavior Modeling in Software and Systems Engineering

Sabah Al-Fedaghi

*salfedaghi@yahoo.com, sabah.alfedaghi@ku.edu.kw*
Computer Engineering Department, Kuwait University, Kuwait

**Summary**
The design of complex man-made systems mostly involves a conceptual modeling phase; therefore, it is important to ensure an appropriate analysis method for these models. A key concept for such analysis is the development of a diagramming technique (e.g., UML) because diagrams can describe entities and processes and emphasize important aspects of the systems being described. The analysis also includes an examination of ontological concepts such as states and events, which are used as a basis for the modeling process. Studying fundamental concepts allows us to understand more deeply the relationship between these concepts and modeling frameworks. In this paper, we critically analyze the classic definition of a *state* utilizing the Thinging machine (TM) model. *States* in state machine diagrams are considered *the* appropriate basis for modeling system behavioral aspects. Despite its wide application in hardware design, the integration of a state machine model into a software system's modeling requirements increased the difficulty of graphical representation (e.g., integration between structural and behavioral diagrams). To understand such a problem, we project (create an equivalent representation of) states in TM machines. As a case study, we re-modeled a state machine of an assembly line system in a TM. Additionally, we added possible triggers (transitions) of the given states to the TM representation. The outcome is a complicated picture of assembly line behavior. Therefore, as an alternative solution, we re-modeled the assembly line based solely on the TM. This new model presents a clear contrast between state-based modeling of assembly line behavior and the TM approach. The TM modeling seems more systematic than its counterpart, the state machine, and its notions are well defined. In a TM, states are just compound events. A model of a more complex system than the one in the assembly line has strengthened such a conclusion.

## 1. Introduction

The design of complex man-made systems mostly involves a modeling phase; therefore, it is important to ensure an appropriate analysis (understand, design, and evaluate) method for these models and their fundamental concepts. The analysis can be viewed as including software engineering and philosophy simultaneously [1].

A key concept for such analysis is the development of a diagramming technique because diagrams can describe entities and processes, provide documentation, communicate ideas, and emphasize important aspects of the systems being described [2]. For example, The Unified Modeling Language (UML) and its profile are considered a suitable specification language for the design of systems.

In this paper, we focus on state machines, which are considered *the* appropriate basis for modeling system behavioral aspects. Besides the state-machine concept, other models have been invented, often inspired by the idea of states (e.g., the models used in PLC languages). The central idea of these models is a state. The state machine stands as the preferred model for describing systems' behavior [3].

### 1.1 State Machines

Finite-state machines (FSMs) are well-established computational abstract devices and are used at the heart of most digital design [4]. FSM models are widely utilized to specify systems in such fields as sequential circuits, distributed systems, communication networks, and communication protocols. They can also be used to model software systems' behavior. A "state machine can solve a large number of problems, among which is electronic design automation, communication protocol design, parsing and other engineering applications. In biology and artificial intelligence research, state machines are sometimes used to describe neurological systems and in linguistics, to describe the grammars of natural languages" [5].

In software engineering, an FSM models the behavior of a single "object," specifying the sequence of "events" that an object goes through during its lifetime. It takes inputs and produces outputs by following a set of rules determined by the internal state of the system. Typically, "behavior" refers to how the software will respond to external events (sometimes called *triggers*). According to Wagner and Wolstenholme [3], "Probably, the state machine is the only known model (of the many used in software development) that really gives a designer a chance to verify a control system and thus, it is the only way to produce reliable control software."

Events trigger transitions between states. A state is established by its relations to other states and to inputs and outputs. A machine is in one indivisible state at a time. The current input plus the current state determine the following output and the machine's next state. FSMs have been extended by developing the so-called statecharts, which provide the possibility to model states at multiple hierarchical levels.

## 1.2 States

A *state* is understood as a static situation, such as waiting for some external event to occur. When a state is entered, it becomes active, and it becomes inactive if it is exited. In this paper, we focus on simple states (i.e., ignoring composite and submachine states). A state is also described as an "abstraction of the values and links of an object" [6]. According to Blaha and Rumbaugh [6], "sets of values and links are grouped together into a state according to the gross behavior of objects." States often correspond to verbs with an "ing" suffix (Waiting, Dialing) or the duration of some condition (Powered, Below Freezing). Also, events represent points in time, and states represent intervals of time; however, according to Blaha and Rumbaugh [6], "Of course, nothing is really instantaneous; an event is simply an occurrence that an application considers atomic and fleeting." On the other hand, some thinkers consider events subtypes of states [7].

## 1.3 Difficulties

Despite its wide application in hardware design, the integration of the state machine model into a software system is accomplished with some "new ideas or reinventions" [3]. Some extensions and changes in the state machine terminology have increased the difficulty of graphical representation of state machines. According to Wagner and Wolstenholme [3], the definition of a finite state machine seems to "require discussion." The concept is still not well understood or interpreted in the software domain despite its broad application in hardware design [3]. Misunderstandings about state machines have produced several stories and half-truths. The concept of the state machine has (unintentionally?) been reinvented for software several times [3]. According to Steward [8], a misunderstanding of the nature of states and of their role in causal explanation has led to a seriously distorted understanding of states. According to Baldawa [9], "We should bear in mind that even though state machines are powerful tools to solve certain kinds of problems, it is not a panacea for all your database modeling problems and not all problems can be modeled using state machines."

In this paper, we try to analyze critically the classic definition of a state in state machines, utilizing the Thinging machine (TM) model [10-11]. Conceptual modeling includes ontological concepts such as states and events that are used as a basis for the modeling process. Analyzing fundamental concepts in conceptual modeling allows us to understand more deeply the relationship between ontological concepts and modeling frameworks.

## 1.3 About this paper

The next section contains a brief description of TM modeling. In Sections 3 and 4, we project the states of an assembly line example in a corresponding TM machine, presenting a clear contrast between state-based behavior modeling and the TM approach. In section 5, we analyze a more complex system of a telephone line given in Blaha and Rumbaugh [6].

## 2  Thinging Machine (TM)

TM views the world as thimacs (things/machines) constructed from nets of subthimacs. Modeling consists of a lower (static) structure of things that are simultaneously machines, and both merge into a thimac. At the upper level (dynamics), a time thimac combines with the static thimac to generate events.

The thimac is an encapsulation of a thing that reflects the unity and hides the thimac's internal structure, and a machine (see Fig. 1) shows the structural components (static: outside of time – called region), including *potential* actions of behavior. The static "thing" does not actually exist, change, or move, but it has potentialities for these actions when combined with time. A TM event is an encapsulation of a region and a time.

A thimac is a *thing*. The thing is what can be created (appear, observed), processed (changed), released, transferred, and/or received. A thing is manifested (can be recognized as a unity) and related to the "sum total" of a thimac. The whole TM occupies a conceptual "space" that forms a network of interrelated thimacs that together form an organic whole.

The thimac forms a compositional structure, in which elementary thimacs combine in systematic ways to create compound new thimacs, allowing us to make infinite thimacs structure. The result is compositional "world" models built to represent things and understand their interactions and relations. The whole is a grand thing/machine. Thimacs can be "connected" only via flow connections among thimacs. Therefore, things are part of the TM static description (static model) and are part of the dynamic model when merged with time to form events.

The thimac is also a *machine* that creates, processes, releases, transfers, and/or receives. Fig. 1 shows a general picture of a machine. The figure indicates five "seeds" of potentialities of dynamism: creation, processing, releasing, transferring, and receiving.

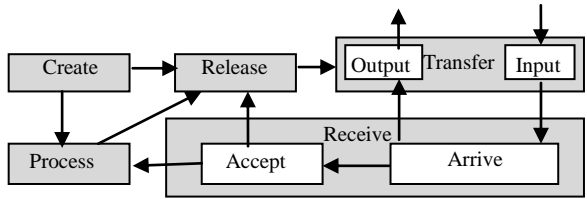

Fig. 1 Thing machine.

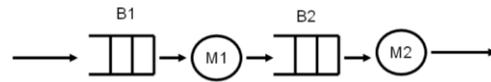

Fig. 2 An example of an assembly line (From [12]).

All things are created, processed, released, transferred, and received, and all machines create, process, release, transfer, and receive other things. Things "flow through" (denoted by a solid arrow in Fig. 1) other machines. The thing in a TM diagram is a presentation of any "existing" (appearing) entity that can be "counted as one" and is coherent as a unity.

Fig. 1 can be described in terms of the following generic (having no more primitive action) actions:

**Arrive**: A thing moves to a machine.

**Accept**: A thing enters the machine. For simplification, we assume that all arriving things are accepted; therefore, we can combine the arrive and accept stages into one stage: the **receive** stage.

**Release**: A thing is ready for transfer outside the machine.

**Process**: A thing is changed, handled, and examined, but no new thing results.

**Create**: A new thing is "coming into being" (found/manifested) in the machine and is realized from the moment it arises (emergence) in a thimac. Things come into being in the model by "being found." The "ceasing to be" of a thing can occur anywhere in the model and can be represented as a freezing storage (graveyard) in the model.

**Transfer**: A thing is input into or output from a machine.

Additionally, the TM model includes the **triggering** mechanism (denoted by a dashed arrow in this article's figures), which initiates a flow from one machine to another. Multiple machines can interact with each other through the movement of things or through triggering. Triggering is a transformation from one series of movements to another.

## 3 State-based Modeling vs. TM Modeling

In this section, we project the states of an assembly line given by Huang [12] (Fig. 2) in a corresponding TM machine. The resultant TM model shows that that the model does not cover many details besides states. We resolve this lack of details by adding possible triggers (transitions) of the given states. The outcome is a complicated picture of behavior. Therefore, we contrast this state-based method with a re-modeling of the assembly line in a TM.

According to Huang [12], a state is a unique status of a system at a particular time. A discrete event system is a system in which the state does not change between consecutive events.

An approach to reduce the number of system states is to describe components' states instead of all of a system's states. Huang [12] gives an example (Fig. 2) of an assembly line that includes two machines, M1 and M2. Each machine has unit capacity and has a preceding buffer with a capacity of three items. The states in this system can be captured in four components (B1, M1, B2, and M2). The set of all possible state values of B1 is (0, 1, 2, or 3). Because the succeeding buffer could block M1, there are three possible state values, (idle, busy, or blocked). M2 has two possible state values, (idle or busy). The number of all possible system states is 96 [12].

### 3.1 TM perspective of the assembly line states

Fig. 3 shows the TM model of this state-based modeling of an assembly line. Items are received in M1 (numbers 1 and 2) to trigger incrementing (3) the number of items in B1. We assume that initially, the number of items in B1 is zero (4). If the number of items in B1 is greater than zero (5), then an item is released (6) from B1 to be processed (7) in M1. Also the number of items in B1 is decremented (8). These processes of incrementing and decrementing trigger the creation (9) of new values for the number of items. When M1 receives an item from B1, it blocks any further release (10) because it will be busy processing the received item.

When M1 releases (11) an item, after processing it, to M2, assuming that M2 is not blocked (12), the blocking of further release from B1 is lifted (13), so it can check whether B1 contains any items (14). When M2 receives the item (15), the number of items in B2 is incremented (16). If the number of items in M2 is greater than zero (17), an item is released from B2 to M2 (18). Also, that number is decremented (19). These operations of incrementing and decrementing trigger the creation of the number of items in B2 (20). When M2 receives an item from B2, it blocks further reception (21). When M2 finishes processing the item (22), the blocking of further reception from B2 is lifted (23).

The situation of whether M2 is blocked with regards to receiving further items from M1 is registered regardless of whether the number of items in B2 reaches 3 (24). This situation is communicated to M1 (25) and affects the decision to send items from M1 to M2 (26 and 13).

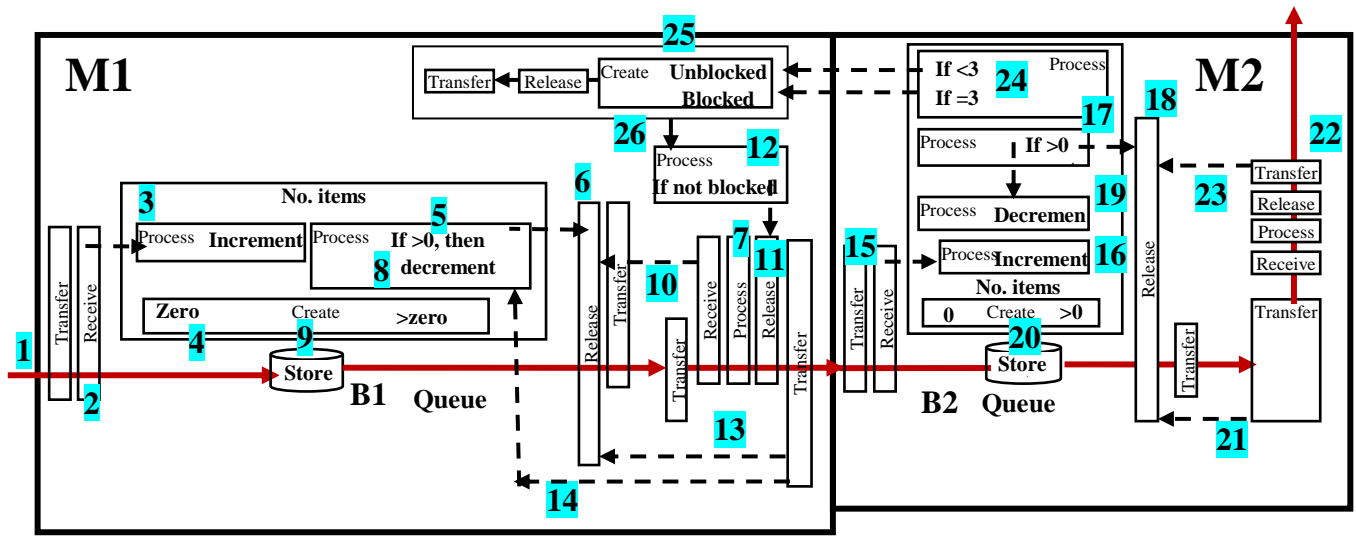

Fig. 3 The TM static model of the assembly line.

## 3.2 States as TM events

An event in a TM is a subdiagram of the static model (called the region of the event) plus a time machine. It can endure, or it may be instantaneous. The five generic actions form five generic events. They may be compound, forming "higher order events" consisting of generic events fitting together as subthimacs to produce larger thimacs. Fig. 4 shows the event *An Item has arrived to M1 and added to B1*. For simplicity, we represent an event by its region.

According to Steward [8], events *happen* whereas states *obtain* (or do not obtain). From the TM point of view, obtaining is achieved as the result of the create event. In TM, the so-called states are treated as TM events. To support this thesis about states (model focus only), in such a discussion, we appeal to an indication of that (states as TM events) in some aspects of "world"-based ontology.

First, we claim that states are results of generic events (generic actions plus time). Therefore, states are compound events of generic events. For example, *temperature and pressure* states in gases are triggered by movements (changes in position – generic actions) of the molecules of the gases. Thus, the events of movements of the molecules trigger the events of changes in temperature and pressure. Both types of events involve dynamism: generic actions of movement and the increase/decrease (process) in values. Even states of *equilibrium* in certain chemical systems might involve the constant passage of molecules between their liquids. Gaseous states [8] involve the generic events (generic actions plus time), create, process, release, transfer, and receive, which implies that the so-called states involve changes.

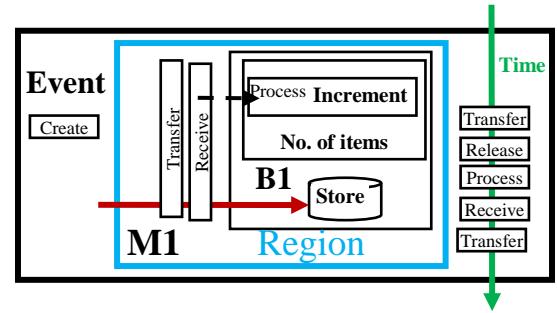

Fig. 4 The event *An Item has arrived to M1 and added to B1*.

Returning to Huang's [12] example (Fig. 3), let us locate the assembly line states in the TM model as events. Fig. 5 shows such a model built, according to Huang [12], of states, using the static model in Fig. 3. Note that in a TM, the so-called objects (continuants) are a type of event (have at least the *create* action) (e.g., 0, 1, 2, and 3 in Fig. 5). Even though "continuants" persist through time and exist as wholes at every moment of their existence (starting from create), in a TM, such an existence is divided into a sequence of events that involves processing, releasing, transferring, and receiving. Analogously, water's being at 90°C, for instance, seems to be a state which exists, as it were, in full at all times at which the water is at that temperature [8]. In a TM, this state is modeled as heat flows (actions) in water, and processing that heat triggers an increase in the water temperature to 90°C. The whole event (being at 90°C) is a composite event of generic events of continuous supplying of heat. The state is incomplete whenever one of the generic events (transfer/receipt of heat) is incomplete, just as a football match is incomplete when some actions in the game stop.

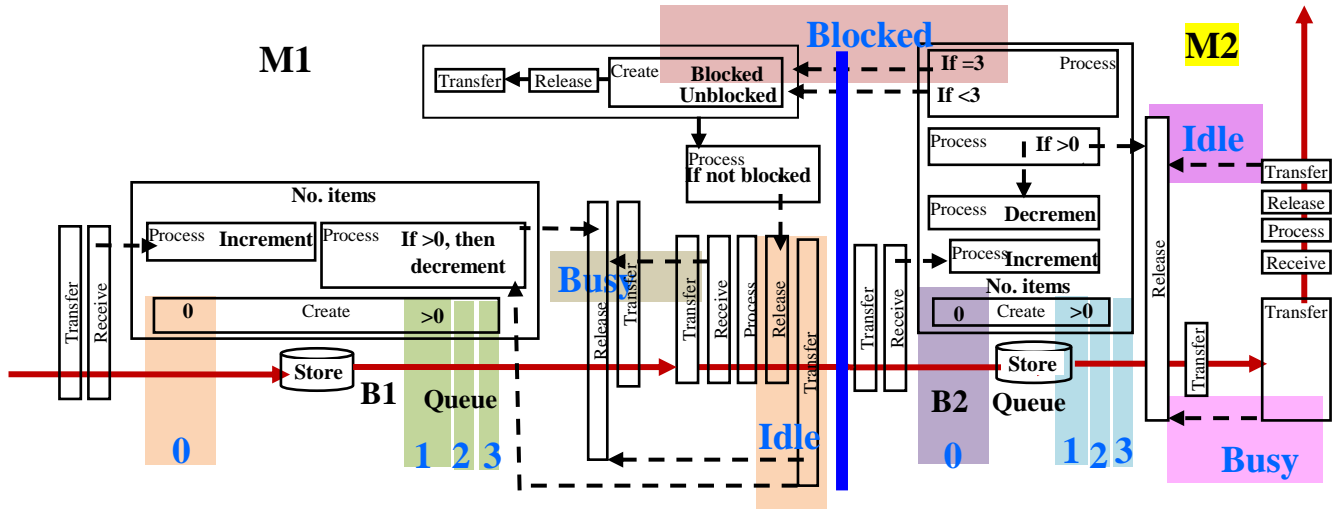

Fig. 5 The given states superimposed on the TM description (for simplicity, we removed some boxes).

To elaborate on such an issue, consider *knowledge* (classical mental *state*) as an event when the knowledge is permanent or long-lasting over a person's life. Fig. 6 shows the TM model of this situation, in which the involved events include receiving information, processing it, storing it as knowledge, and processing stored knowledge to be released. In a TM, creating knowledge is the event of *starting the existence* of the knowledge as a thing in the memory. This *existence* reflects knowledge as *a being in the model*, in contrast to the general notion of existence in the "world." It is an event (occurrence) in the sense that it is an announcement of the availability of knowledge to participate in the dynamism of the world through other events. It is an event that creates a thing.

The *knowledge*, after its creation, becomes a *thing*. This thing is now not a continuous creation of knowledge but a finished event. After the creation, the event-ization finishes and *thingness* starts. Therefore, knowledge is not a lifelong event; rather, it is a thing that participates in many events whose regions include knowledge. The creation event of a thing is a unique event that produces a thing that "is there"; however, create-ness has finished as an event. A thing "being there" is not a long-lasting event; rather, the thing continued *being* inside other events.

Suppose that a person is born (created) but never participates or appears in any further event (in the model). We claim that this is not possible because a TM's existence implies eventi-zation. There is not a second in a person's life that does not involve participating in some events (e.g., after birth, sucking milk, growing, crying). Suppose that a thing, E1, is created, but it was not involved in any event for a certain period, and then it appears in an event, E2. We claim that in this "disappearance" period, the time slice between E1 and E2, contains all the events that are not detected because they are not of interest/capability to/of the modeler (observer).

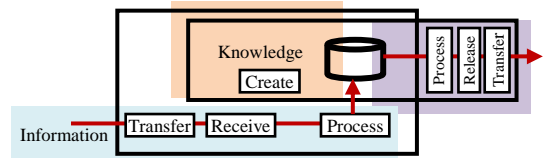

Fig. 6 Events of getting information, processing it, storing it as knowledge, and processing stored knowledge to be released.

TM existence of a thing after the creation event is its participation in a sequence of events.

Consider a very static object, such as Cleopatra's Needle on the Victoria Embankment in London. According to Whitehead [13] Cleopatra's Needle "is a series of events. It is actively happening. It never remains the same. A physicist who looks at that part of the life of nature as a dance of electrons will tell you that it loses molecules and gains others daily. Even the plain man can see that it gets dirtier and is occasionally washed" [13]. Even knowledge in the mental domain, after it is acquired, goes through continuous mental events that diminish, update, or modify it.

Accordingly, we claim that there is a unique type of event, *create*, that results in existence. We are developing this thesis in the context of modeling where existence refers to being (appearance) in the model. Therefore, a TM's existence begs the question "What is modeled?" A dead person (a thing) can exist in a TM if it participates in an event such as, say, celebrating his/her memory. Also, a round square "exists" if it is created (appears) in the TM description (e.g., it is talked about). For example, *There being round squares is impossible* [7] is modeled as the creation of a box for impossibility and, inside it, creating a box for a round square. To make such a subdiagram an event, it is complemented with a thimac of "all times" (create and process time).

Additionally, such a state (of an affair) as *Abdul-Jabbar's being more than seven feet tall is* a creation event that occurs at a certain point in Abdul-Jabbar's life. Many events occur for him during his life. However, these events are not things that hang from him as leaves attached to a tree; rather, whenever his height participates in a future event, it flows (transfer, receive) from the creation event to the new event. The new event "constructs" Abdul-Jabbar by importing continuing "attributes" from previous events.

We will apply this claim that the *state* is a type of *event* in Huang's [12] assembly line example.

### 3.3 Projection States as events

Returning to Fig. 5, which shows states projected [12] over the TM model, we note that the diagram is missing many TM details. In a state machine, more details can be specified as triggers of the states. Fig. 7 identifies possible triggers of states. The resultant diagram of the whole approach of identifying states and triggers seems to produce an unsystematic way to specify the assembly line's behavior. It seems that there is no reason for this top-down process that starts from compound events given as states and triggers. A simpler approach is to construct the TM diagram from scratch, identify a suitable set of events, and specify the behavior as a chronology of these events.

To illustrate such an approach, consider the simple state machine of a door given by Sparx [14] and shown in Fig. 8. First, the door is described in terms of a situation (state) with an initial state and create trigger. Then *Close/[doorWay->is Empty]* and *Open/* are used to described the changes in the door's position. Alternatively, Fig. 9 shows the TM static description of a door.

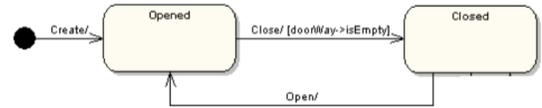

Fig. 8 An example of a state machine (partial, from [14]).

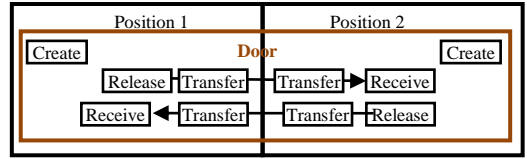

Fig. 9 Static TM model of the door.

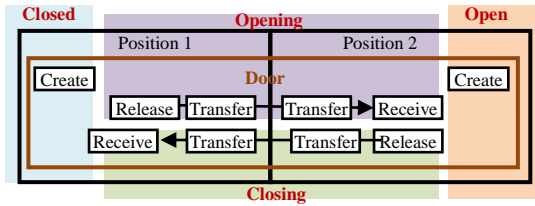

Fig. 10 Dynamic TM model of the door.

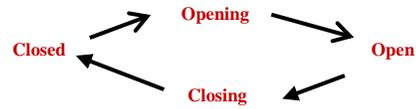

Fig. 11 The behavioral model.

The diagram reflects the intuitive idea of events as the door moves between two positions; therefore, as Fig. 10 shows, there are two events, closed and open, and two changing events, opening and closing. Fig. 11 shows the corresponding behavioral model. The TM model seems simpler than the state diagram, but it is richer in semantics.

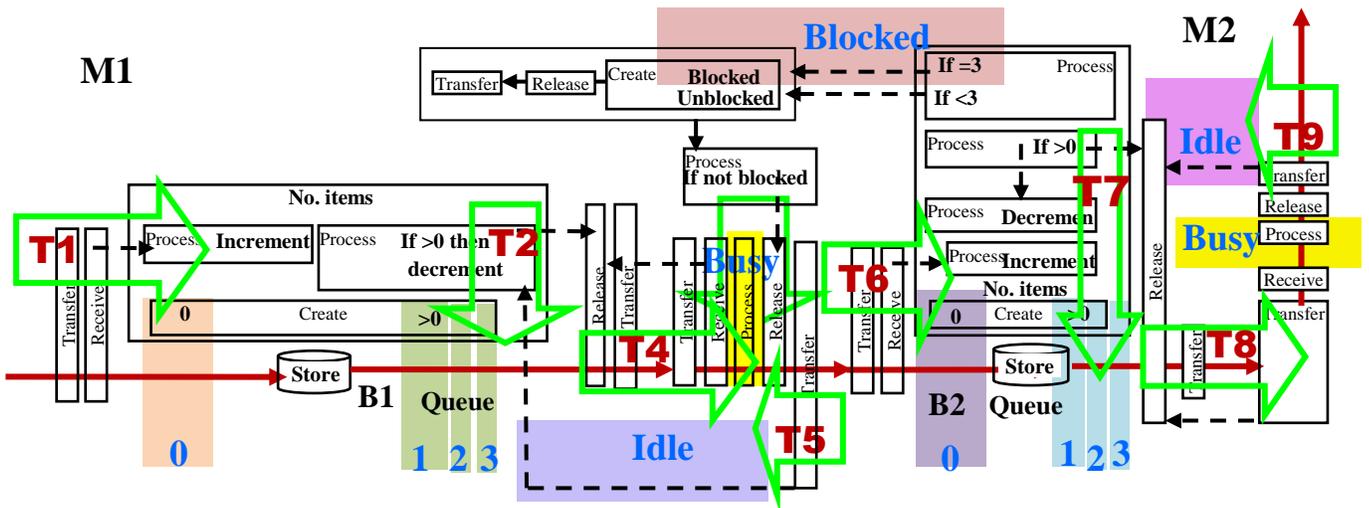

Fig. 7 The given states with triggers.

## 4    TM model of the assembly line

### 4.1 The static model

Fig. 12 shows a fresh start to model the assembly line example. First, an item is received in M1 (number 1). The number of items in B1 is incremented (2). The number of items in B1 is checked.

- If it is 3, then no additional item is accepted in M1 (3).
- If it is greater than 0 (4), then an item is released to M1 (this happens automatically because M1 is initially idle; later, M1 will trigger this action), the number of items in the queue is decremented (5), and accepting an item from the outside is activated (6).

Accordingly, an item moves from B1 to M1 (7 and 8) and processed (9). If M1 is not blocked, (10) flows to M2 (11). The number of items in B2 is incremented (12). If the number of items in B2 is 3, then M2 is blocked (13); otherwise, it is unblocked. If the number of items in B2 is >0 (14), then an item is sent to M2 (15) and the number of items in B2 is decremented (16). The received item is processed and leaves M2 (17).

### 4.2 The Events model

Accordingly, Fig. 13 shows the set of events superimposed on the static model as follows:

E1: A new item enters M1, and the number of items is incremented in B1.

E2: The number of items in B1 reaches 3, so further acceptance of items is blocked.

E2: If the number of items in B1 is greater than zero, send an item to M1.

E4: Decrement the number of items and accept further items in B1.

E5: An item flows from B1 to M1.

E6: An item is processed in M1.

E7: M2 is not blocked.

E8: If M2 is not blocked, then send the processed item from M1 to M2.

E9: Increment the number of items in B2.

E10: If the number of items in B2 is 3, then block M2 from accepting an item from M1; otherwise, unblock it.

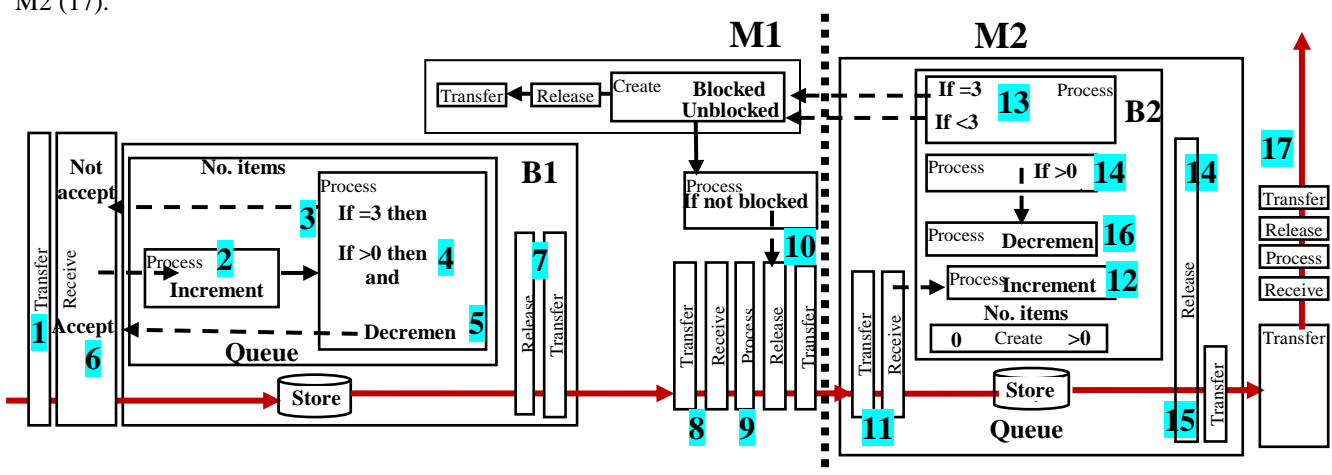

Fig. 12 The TM static model of the assembly line.

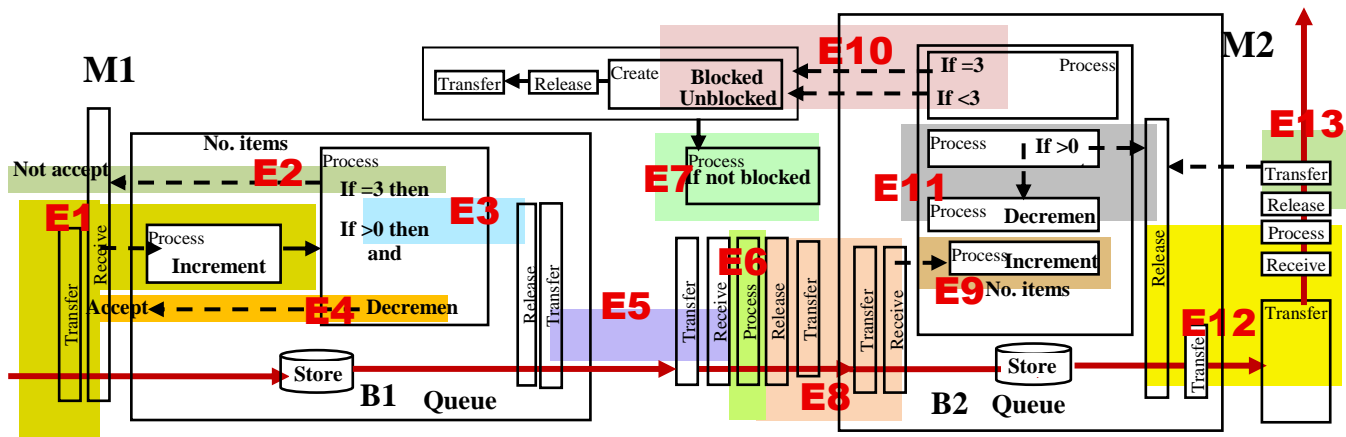

Fig. 13 The TM events model of the assembly line.

E11: If the number of items in B2 is greater than zero, then decrement the number of items.

E12: Send an item from B2 to M2.

E13: Process an item in T2 and release it when finished processing.

Fig. 14 shows the behavioral model of the assembly line. The model represents the chronology of events and stands as the controlling module of the assembly line system.

Suppose that S1 stands for an input item to the assembly line. Fig. 15 shows a sample sequence of events, in which

- it is assumed, initially, that B1 and B2 contain zero items and M1 and M2 are unblocked and idle,
- bookkeeping events (e.g., E3 and E4) are shown only when they are applicable,
- all events have the same time duration, and
- we assume priority of events from left to right.

## 5. Another example of TM Modeling

We noted previously that that the TM model includes more details than are necessary to produce an accurate basic model. Of course, in general, such details come with the cost of having a more complex model system. However, we think that TM modeling has appropriate level of detail because it is constructed from one category: the thimac. The thimac includes five actions, flows and trigger that are applied repeatedly while a high level abstraction is preserved. Additionally, the TM modeling seems more complete than a state machine because it starts with a static model, then includes an events model, and last, the behavioral model

To emphasize such a thesis, we analyze in this section a more complex system than the one in the previous section.

Consider the state machine given in the classical book *Object-Oriented Modeling and Design with UML* [6] in Fig. 16. The state diagram is for a telephone line. According to Blaha and Rumbaugh [6], the diagram concerns a phone line and not the caller or the callee. It contains sequences associated with normal calls as well as some abnormal sequences, such as timing out while dialing or getting a busy signal.

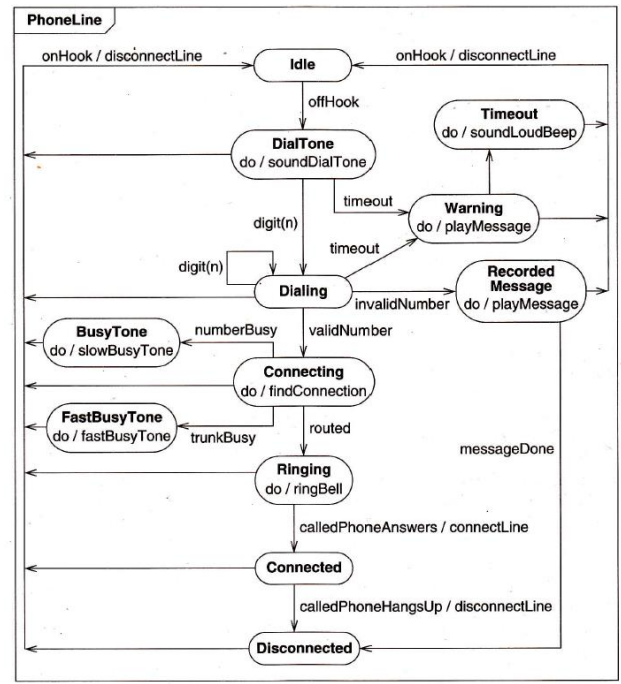

Fig. 16 State diagram for phone line with activities (From [6]).

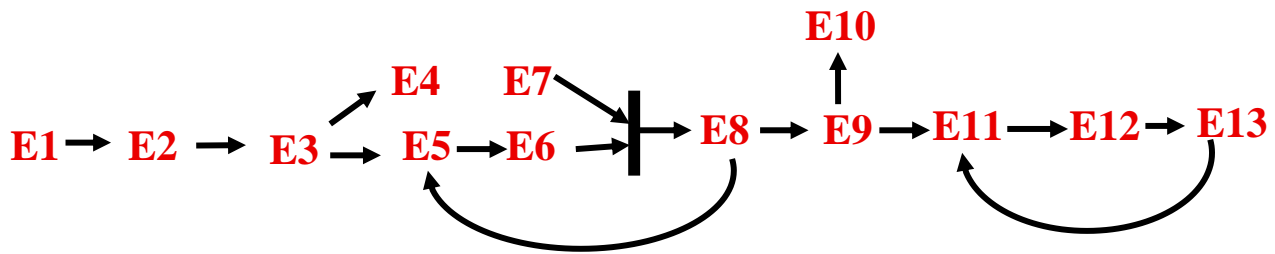

Fig. 14 The TM behavioral model of the assembly line.

| Time 1 | Time 2 | Time 3 | Time 4 | Time 5 | Time 6 | Time 7 | Time 8 | Time 9 | Time 19 |
|--------|--------|--------|--------|--------|--------|--------|--------|--------|---------|
| E1: S1 arrives to M1 | E1: S2 arrives to M1 | E1: S3 arrives to M1 | E1: S4 arrives to M1 | E2: Block arrivals to M1 | E4: Unblock arrivals to M1 | E5: S2 is sent from B1 to M1 | E6: S2 is processed | E8: S2 is sent to from M1 to M2 | E12: S2 is sent from B2 to M2 |
| | E5: S1 is sent from B1 to M1 | E6: S1 is processed | E8: S1 is sent to from M1 to M2 | E12: S1 is sent from B2 to M2 | E13: S1 leaves M2 | E1: S5 arrives to M1 | E2: Block arrivals to M1 | | E4: Unblock arrivals to M1 |

Fig. 15 Sample partial run of the behavioral model.

According to Blaha and Rumbaugh [6], "At the start of a call, the telephone line is idle. When the phone is removed from the hook, it emits a dial tone and can accept the dialing of digits. Upon entry of a valid number, the phone system tries to connect the call and route it to the proper destination. The connection can fail if the number or trunk are busy. If the connection is successful, the called phone begins ringing. If the called party answers the phone, a conversation can occur. When the called party hangs up, the phone disconnects and reverts to idle when put on hook again" [6].

## 5.1 TM Static Model

Fig. 17 shows the TM model of this phone line. When the phone is lifted from the hook (1), a signal is created (2) and omitted (for simplification, we do not show the flow of this signal).

This signal triggers the following initialization;
- A dial tone (3) that indicates acceptance of the dialing of digits.
- The number of digits is initializing to zero (4).
- The dialed number is initialized to blank (5).
- Timing is set on (6).

Accordingly, with each created digit (7) by the user,
- the number of dialed digits is incremented (8),
- the timing is reset (9), and
- the input digit to construct the dialed number (10) is sent.

The number of dialed digits is examined (11), and
- if it is less than n, then this triggers expecting additional digit input (13);
- if it is = n, then the input number is processed (14), and if it is valid (15), a connection is requested from the other side (16).

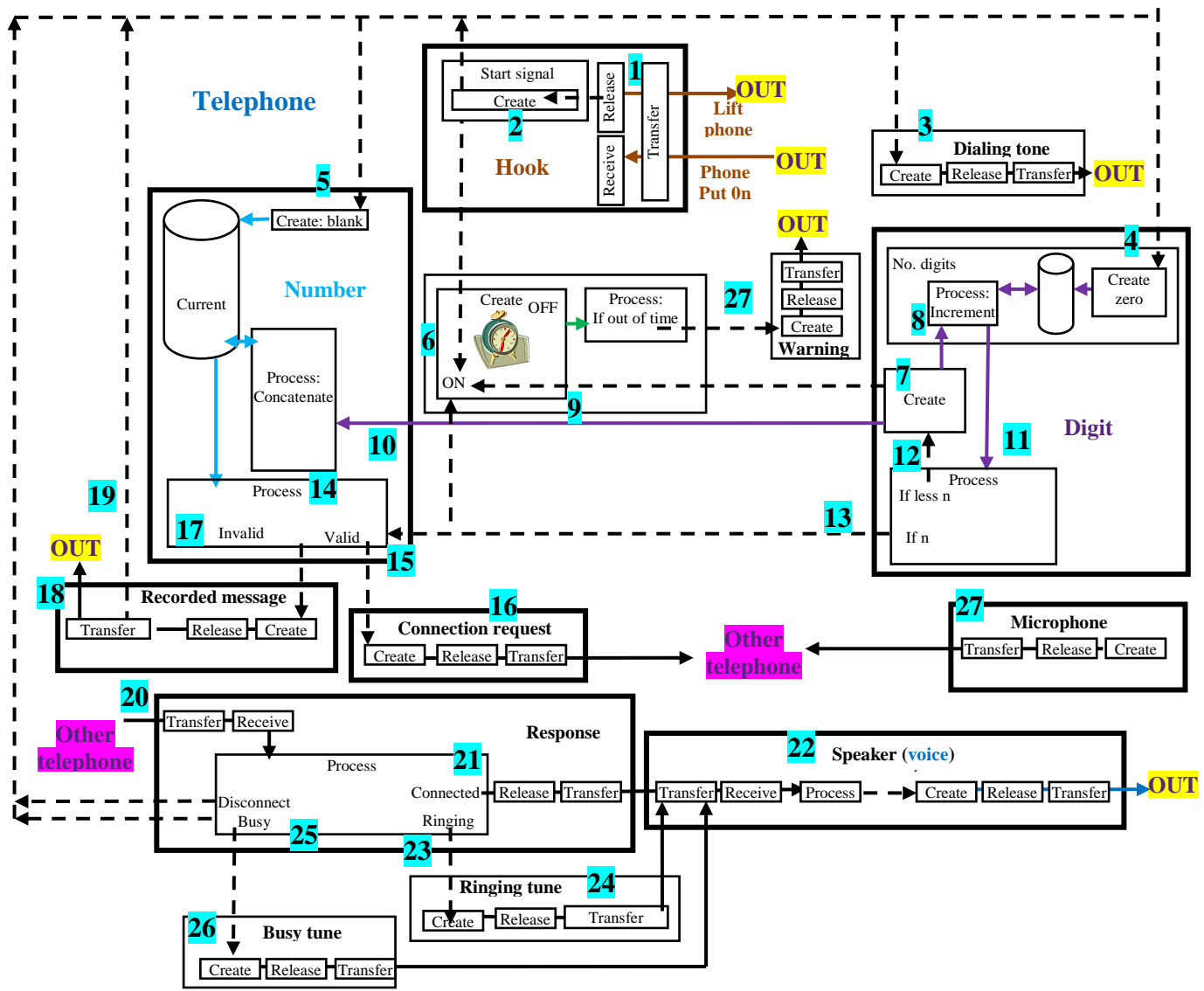

Fig. 17 The static TM representation of the phone line.

- If the number is not valid (17), a recorded message is output (18) and the system is initialized again to accept the dialing of digits (see initialization above).

The other side of the line sends a response (20).

- If it is a connected response (21), it is sent to the speaker (22).
- If it is a ringing tone (23), a ringing tone is produced on the speaker (24).
- If it is busy or disconnected, a busy sound (25) is produced and sent to the speaker (26).

Meanwhile, during the connection, the user sound is captured by the microphone and sent to the other side. Additionally, when time expires (27), a warning time is output.

## 5.2 Projecting States over the TM model

If we project Blaha and Rumbaugh's [6] states over the TM representation of the phone line, we end up with many details the state diagram does not include. To save space, Fig. 18 shows projecting the state of a dial tone, which includes many initializations when the phone is removed from the hook. Applying triggers would produce very a messy diagram.

## 5.3 Events model

Fig. 19 shows the events model of the phone line. To save space, we will not list these events here. Some of Blaha and Rumbaugh's [6] states coincide with the selected events in Fig. 19. Fig. 20 shows the behavioral model.

## 6. Conclusion

We have critically analyzed the notion of a state in conceptual modeling utilizing the thinging machine model. In a TM, a state is defined as a type of an event. In two case studies, we contrast the state-based method with the corresponding TM model. The TM model is more complete than the state-based model because it starts with a static model, followed by an events model to construct a behavioral model. The TM notion of events is well defined in comparison to the state-based model because

- things/machines form thimacs;
- machines have the five-action structure;
- machines create, process, release, transfer, and receive things;
- things are what can be created, processed, released, transferred, and/or received;
- things denote the wholeness of the thimac;
- events are static thimacs with time;
- behavior is the chronology of events; and
- a single static TM model leads to an event model, which in turn results in a behavioral model using the same modeling notations.

Contrasting this with the numerous notations of state charts and other UML heterogeneous multigraphs, we observe that integrating state machines with other types of UML diagrams is difficult.

Nevertheless, this paper is a work in progress that requires further investigation, specifically experimentation with more state machines of various types. This investigation could also be supplemented with philosophical issues related to states, events, facts, and propositions, which we will do in future research.

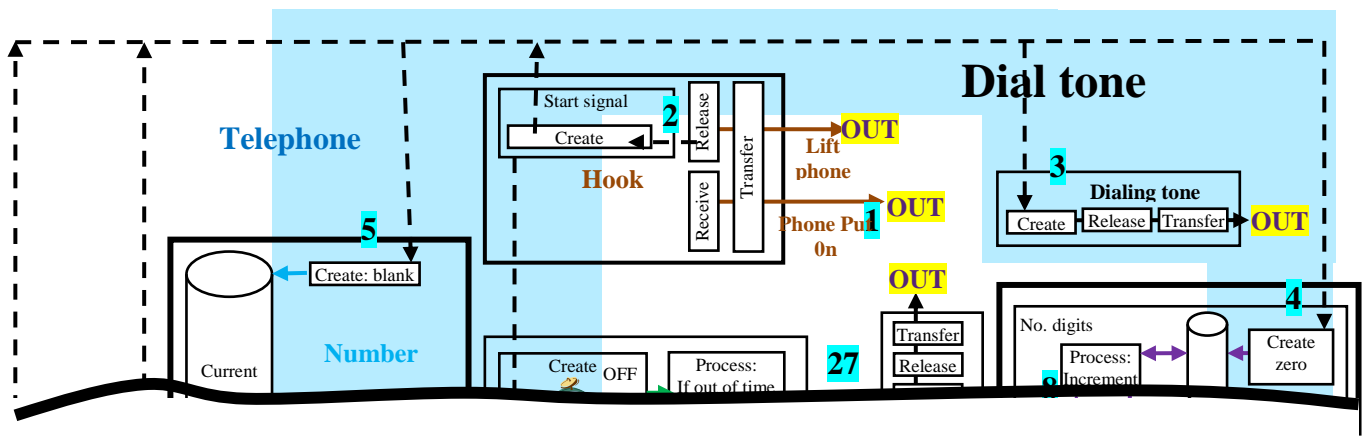

Fig. 18 Partial view of the static model of the phone line showing the part that represents the state Dial tone.

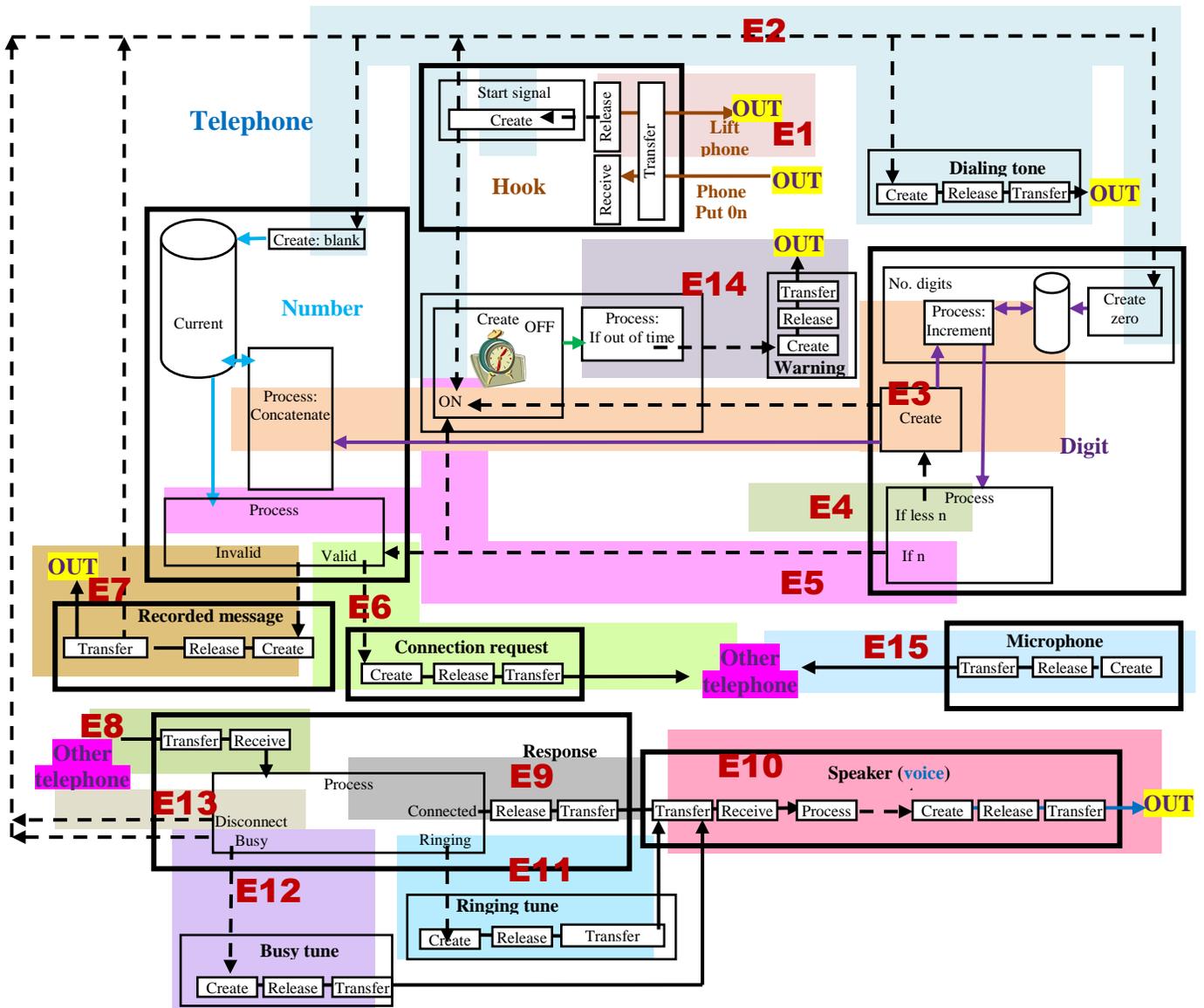

Fig. 19 The events TM representation of the phone line.

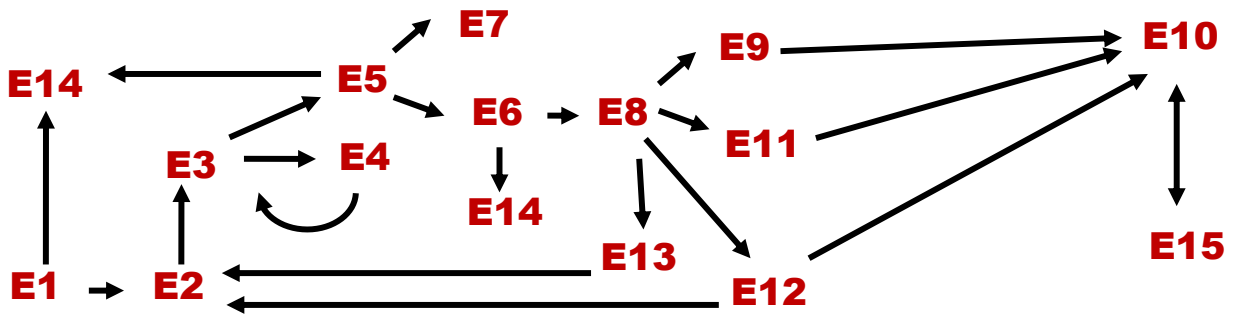

Fig. 20 The behavioral TM representation of the phone line.

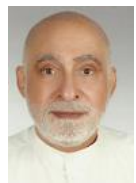

**Sabah S. Al-Fedaghi** is an associate professor in the Department of Computer Engineering at Kuwait University. He holds an MS and a PhD from the Department of Electrical Engineering and Computer Science, Northwestern University, Evanston, Illinois, and a BS from Arizona State University. He has published many journal articles and papers in conferences on software engineering, database systems, information ethics, privacy, and security. He headed the Electrical and Computer Engineering Department (1991–1994) and the Computer Engineering Department (2000–2007). He previously worked as a programmer at the Kuwait Oil Company. Dr. Al-Fedaghi has retired from the services of Kuwait University on June 2021. He is currently (Fall 2021/2022) seconded to teach in the department of computer engineering, Kuwait University.